\documentclass[conference]{IEEEtran}

\usepackage{cite}
\usepackage{color}
\usepackage[stable]{footmisc}
\usepackage{siunitx}
\usepackage{authblk}
\usepackage[center]{subfigure}
\usepackage{verbatim}
\usepackage[super]{nth}
\usepackage{comment}
\usepackage{afterpage,lipsum}

\usepackage{amsfonts}
\usepackage[colorinlistoftodos]{todonotes}
\usepackage[linesnumbered, lined, commentsnumbered, ruled,vlined, longend, noend]{algorithm2e}
\usepackage{algpseudocode}
\usepackage{mathtools}
\usepackage{dirtytalk}
\usepackage{float}
\usepackage{times,amsmath,epsfig,amssymb,cite,bm,float,epstopdf,graphicx,graphics,amsthm}
\usepackage{forloop}
\usepackage{enumitem}
\newcounter{ct}

\newcommand{\cmmnt}[1]{\ignorespaces}

\usepackage{fancyhdr,lipsum}
\fancypagestyle{mahmood}{%
   \fancyhf{} 
   
   \fancyhead[C]{This paper has been accepted for publication in Globecom 2024. You can use this material personally. Reprinting or republishing this material for the purpose of advertising or promotion, creating new collective works, reselling or redistributing to servers or lists, or using any copyrighted component in other works must adhere to IEEE policy. The DOI will be supplied as soon as it becomes available.}

}%
\makeatletter
\let\ps@IEEEtitlepagestyle\ps@mahmood
\makeatother
\begin{document}
\title{External Memories of PDP Switches for In-Network Implementable Functions Placement: Deep Learning Based Reconfiguration of SFCs}

\author{Somayeh Kianpisheh\textsuperscript{1}}
\author{Tarik Taleb\textsuperscript{2}}
\vspace{-4cm}
\affil{\textsuperscript{1}\textit{Centre for Wireless Communications, University of Oulu, Finland}, \textsuperscript{2} \textit{Ruhr University Bochum, Germany} \\
\textit{Emails: somayeh.kianpisheh@oulu.fi, tarik.taleb@rub.de}}
\vspace{-4cm}
\maketitle
\begin{abstract}
Network function virtualization leverages programmable data plane switches to deploy in-network implementable functions, to improve QoS. The memories of switches can be extended through remote direct memory access to access external memories. This paper exploits the switches external memories to place VNFs at time intervals with ultra-low latency and high bandwidth demands. The reconfiguration decision is modeled as an optimization to minimize the deployment and reconfiguration cost, while meeting the SFCs deadlines. A DRL based method is proposed to reconfigure service chains adoptable with dynamic network and traffic characteristics. To deal with slow convergence due to the complexity of deployment scenarios, static and dynamic filters are used in policy networks construction to diminish unfeasible placement exploration. Results illustrate improvement in convergence, acceptance ratio and cost. 
\end{abstract}

\vspace{-0.2cm}
\begin{IEEEkeywords}
In-network computing, Programmable data plane, Network function virtualization, Service function chain
\end{IEEEkeywords}

\IEEEpeerreviewmaketitle

\vspace{-0.6cm}
\section{Introduction}
By In-Network Computing (INC) network devices are programmed to perform computing tasks, in parallel with packet switching; thereby reducing the processing latency and the backhaul traffic \cite{kianpisheh2022survey}. \cmmnt{Network Function Virtualization (NFV) enables the deployment of Virtual Network Functions (VNFs) on virtual machines to enhance the flexibility and reduce the cost, in comparison with traditional special-purpose middle-boxes\cite{kaur2020comprehensive}.} \cmmnt{Most of VNF deployment strategies, as surveyed in \cite{kaur2020comprehensive} consider server/VM-based processing.} INC has been leveraged to deploy VNFs on FPGAs \cite{lopes2021vnfaccel}, Programmable Data Plane (PDP) switches \cite{moro2020framework}, \cite{li2022qos}, smart NICs \cite{dong2020application}, \cite{dong2019application}. This trend has come up to heterogeneous Network Function Virtualization (NFV) environment with hardware-based network devices and Virtual Machines (VMs), to enhance QoS meeting and the cost efficiency. 

Service Function chain (SFC) embedding over heterogeneous NFV environment has been investigated in \cite{sun2017hyper}, \cite{dong2019application}, \cite{dong2020application}. While \cite{sun2017hyper} focuses mostly on implementation of VNFs over network elements, particularly TCAM of switches, studies in \cite{dong2019application}, \cite{dong2020application} provide optimization for SFC embedding with possible deployment of VNFs in VMs and smart NICs. The study in \cite{li2022qos} models the reconfiguration of VNF service trees deployed over VMs and PDP switches, as an optimization to minimize the resource usage and the reconfiguration cost. The study in \cite{xue2021upgrade} provides a method to upgrade the NFV substrate nodes with INC including VNF processing in PDP switches, to maximize the SFCs QoS meeting while maintaining an upgrade budget.

The memory extension of PDP switches has been studied in \cite{xue2020hybrid}, \cite{kim2018generic}. Fig. 1 illustrates an architecture, derived from \cite{xue2020hybrid}, \cite{kim2018generic}. External Memory Access Controller (EMAC) controls the external memory access scheme. The processing of packets are performed in switch pipeline, using the flow tables, PHVs and registers. The space scarcity of TCAM/SRAM can be addressed by remotely placing and accessing flow entries in the external memories (on rack servers). Low-latency, table lookup operations through Remote Direct Memory Access (RDMA) can realize the remote access. Upon packets arrival, in the case that the flow entries are in local memory, they will be processed normally within the pipelines. Otherwise, the miss is escalated to the EMAC. The external memory is read with RDMA and the flow entries are fetched and the actions are performed, leveraging RDMA table. P4 standards and extern functions enable the capability of programming switches for Network Functions (NF) e.g., stream transcoding \cite{aghaaliakbari2023architecture}, firewall \cite{voros2016security}, data encryption/decryption \cite{qin2020flexible} (See \cite{kianpisheh2022survey} for in-network implementable functions). Like \cite{sun2017hyper}, \cite{dong2019application}, \cite{dong2020application}, \cite{li2022qos}, \cite{xue2021upgrade} we focus on SFCs with in-network implementable functions.  

To the best of our knowledge, exploiting external memories for VNF placement in service composition has not been studied. This paper proposes a \textbf{S}ervice \textbf{R}econfiguration method for NFV environment enhanced with \textbf{E}xternal \textbf{M}emories for PDP switches i.e., SR-EM. To avoid detouring of traffic and INC at intervals with ultra-high bandwidth demands, the external memories of switches are exploited to place VNFs that can not be hosted in the local memories due to the large size. Optimization framework is proposed to reconfigure SFCs with minimum deployment and reconfiguration cost, and adaptable with variation of bandwidth and latency demands. 

There is a gap in the literature, to study INC for SFCs under dynamic chains' traffic. Also, the external memory incorporation involves new placement decision variables, SDN controlling/programming, RDMA delay, cost/latency trade-off in the optimization, as well as various deployment scenarios in programming delay and cost calculations. To adopt in-network and server-based processing of VNFs with dynamic network and traffic characteristics, a Deep Reinforcement Learning (DRL) based method is proposed. To cope with complexity of the deployment scenarios in optimization, static and dynamic filters are applied in constructing policy networks. Results show the efficiency in convergence, acceptance ratio and cost.
\vspace{-0.2cm}
\section{Motivation}
\begin{figure*}[!h]
\begin{center}
\begin{minipage}{3.6cm}
\begin{center}
\includegraphics[width=3.6cm, height=3.8cm]{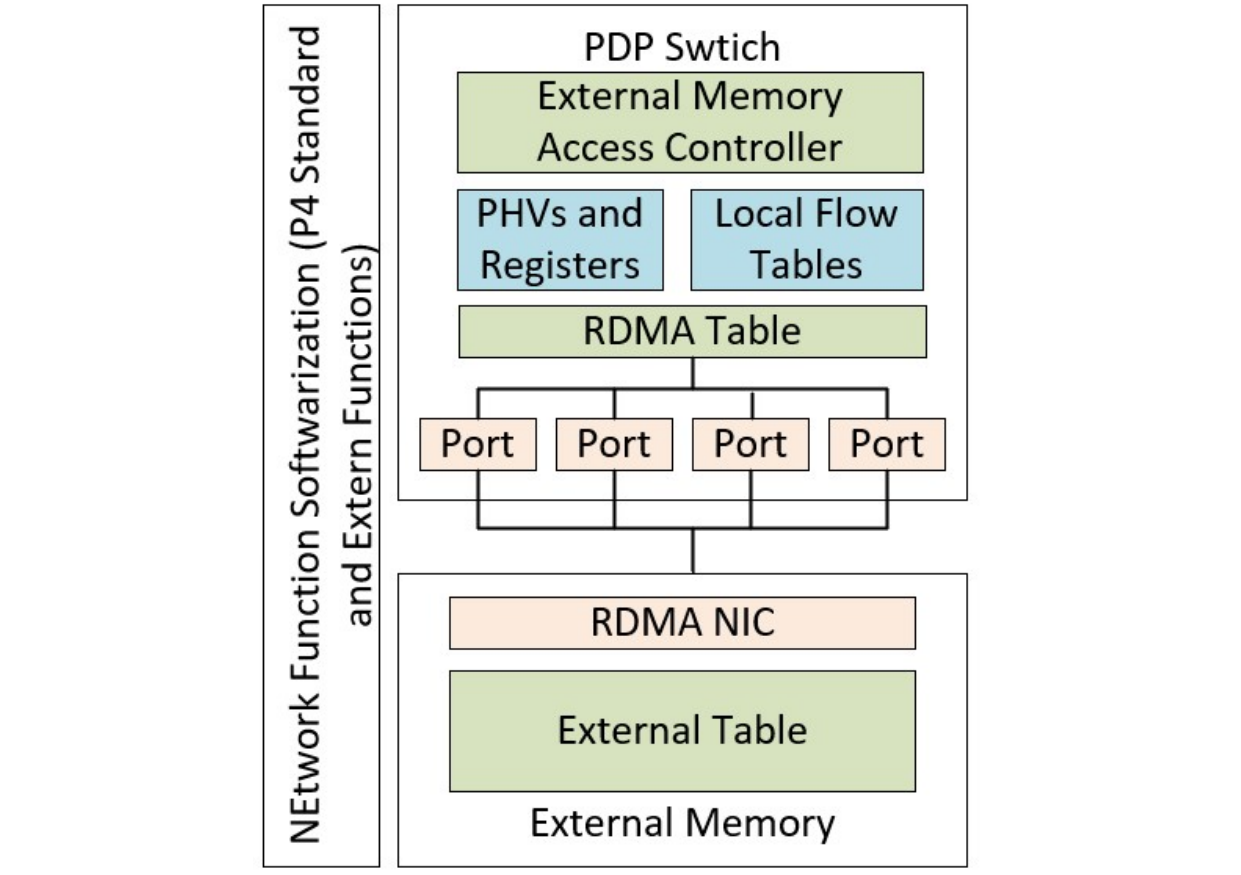}
\label{fig:Switch}
\vspace{-0.5cm}
\caption{The local and external memory in PDP switches.}
\end{center}
\end{minipage}
\begin{minipage}{13cm}
\subfigure(a){
\includegraphics[width=6.3cm, height=3.8cm]{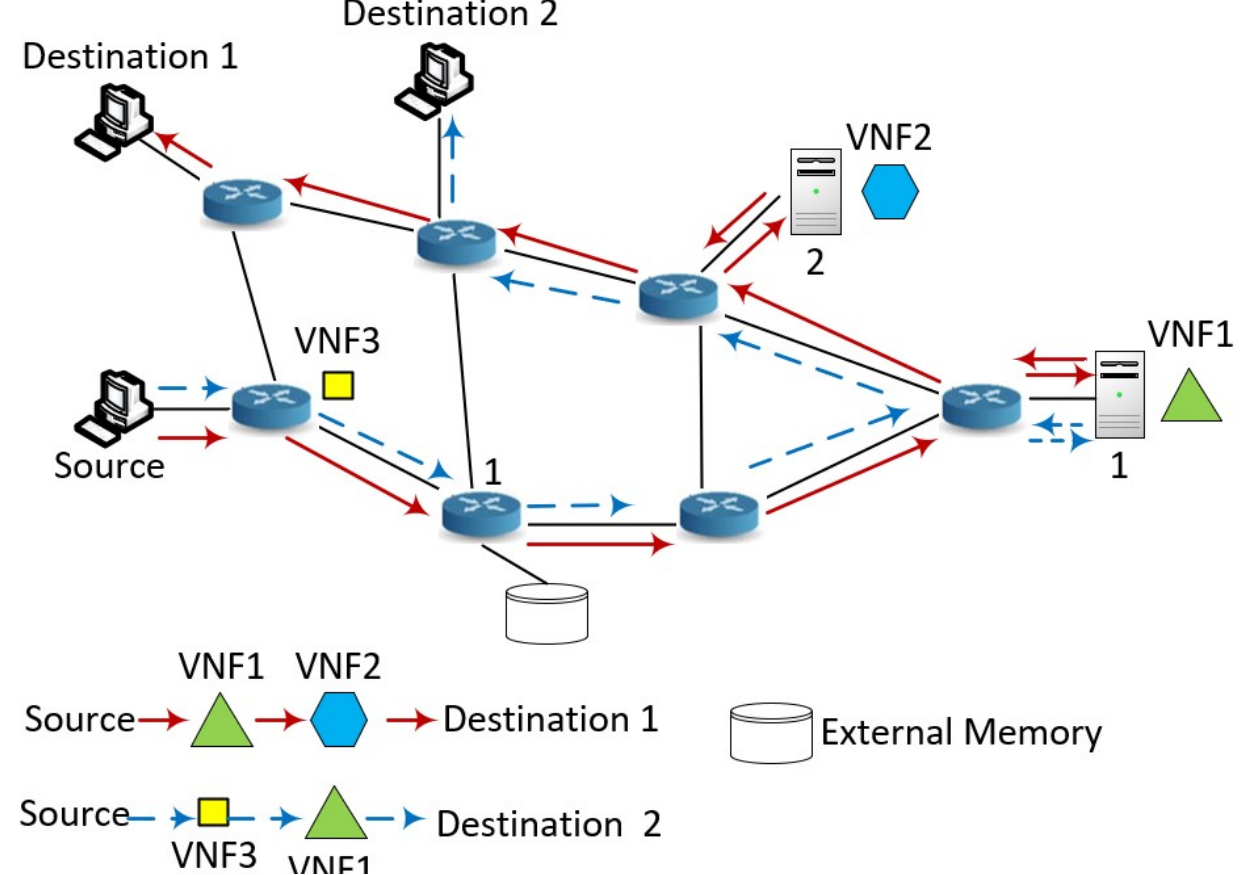}
\label{fig:conf1}
}
\subfigure(b){
\includegraphics[width=6.3cm, height=3.8cm]{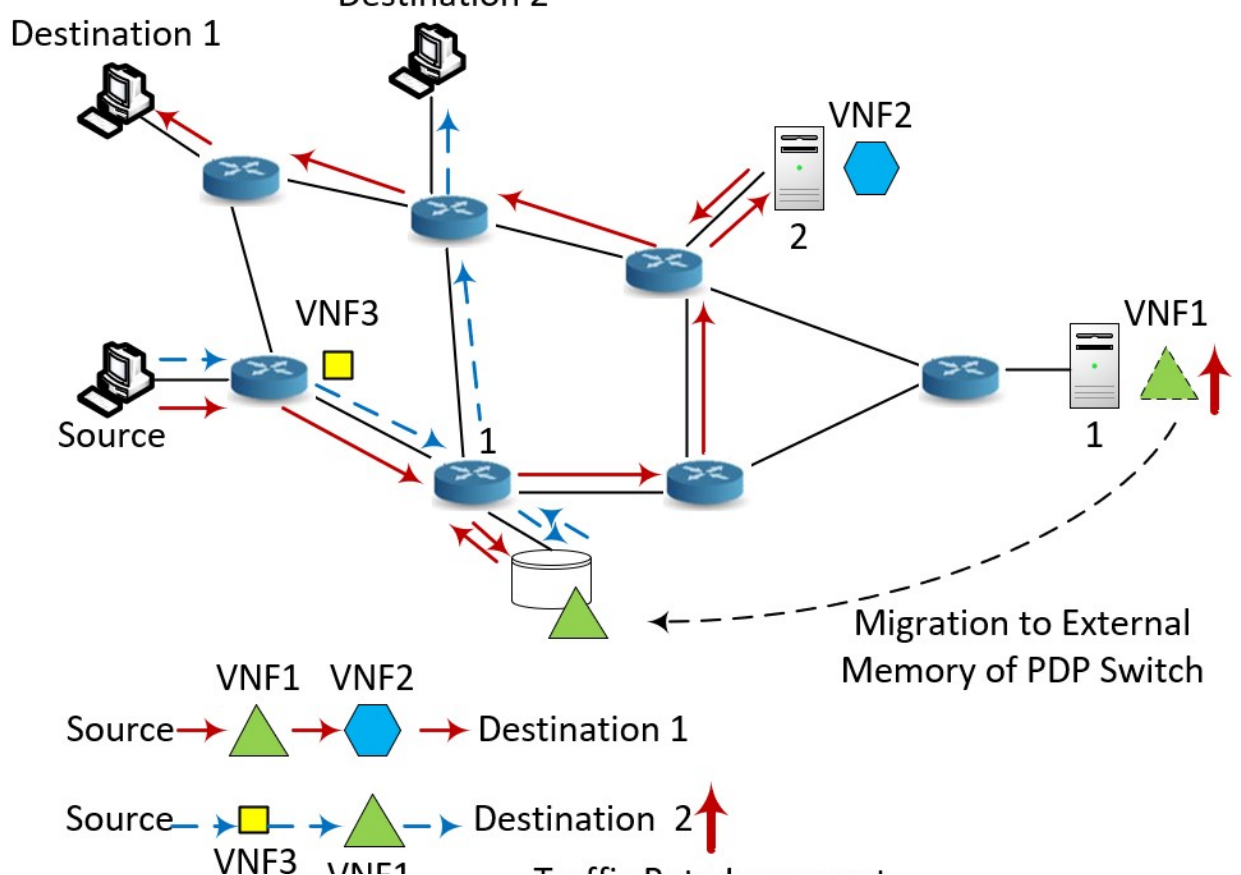}
\label{fig:conf2}
}
\vspace{-0.5cm}
\caption{The scenario for external memory usage. (a) The current configuration with two SFCs. (b) The new configuration when the traffic rate of the second SFC increases so that VNF1 can not accommodate the traffic.} 
\end{minipage}
\end{center}
\vspace{-0.9cm}
\end{figure*}

Let consider a scenario with VNF types: VNF1, VNF2, VNF3. VNF3 is small and can be hosted at any resource. VNF1 and VNF2 are large sizes and can be located on the servers or external memories of switches. There are two SFCs with the origin at source and destinations of 1 and 2: \{VNF1, VNF2\}, \{VNF3, VNF1\}. Fig. 2.a shows an initial configuration at which VNF 3 is processed in the network and prior to the servers, and VNF1 and VNF2 are processed in Servers 1 and 2, respectively. At a time step, the traffic in the second SFC increases, such that the capacity of VNF1 can not accommodate the new traffic, due to server processing limitation. Fig. 2.b, is a new configuration. The processing demand of VNF1 can be provided by deploying it on the switches, however due to its large size it only can be hosted on external memories. Through programming the switch 1, VNF1 is migrated to its external memory. The traffic of both chains can be accommodated due to high processing capability of INC. Avoiding detouring of the traffic to the servers, makes the routing paths shorter and saves the bandwidth consumption.           
\vspace{-0.2cm}
\section{SYSTEM MODEL}
The set of VNF types is $F$ and there is $I_f$ instances for VNF type $f$ with possibility of sharing among SFCs. When $f$ is implemented in the network and as a VM on a server, respectively it consumes $\phi_f^{s}$ (e.g., match-action entries), and $\phi_f^{m}$ storage. The traffic processing capacity of VNF type $f$ when deployed on a switch and server are respectively $C_f^{s}$ and $C_f^{m}$. $Q=\{q_1, q_2 ... q_{|Q|}\}$ is the set of service requests. Request $q$ is modeled as a SFC $G_q = (F_q,L_q, \lambda_q(t), o_q, t_q, d_q(t))$. $F_q$ and $L_q$ are the VNFs and virtual links connecting them respectively. $\lambda_q(t)$ is the SFC traffic rate from a source node $o_q$ to a destination node $t_q$ at time step $t$. The request deadline at time step $t$ is $d_q(t)$. The traffic rate and deadline varies in time slots. 

$G=(V \cup \{v_{c}\},E)$ is the Substrate Network (SN). $v_{c}$ is the SDN controller and $V$ is the set of computing resources including programmable switches $S$ and servers $M$. $E$ shows connectivity with entry $e_{u,v}\in{E}$ to be 1 when node $u$ is directly connected to node $v$, with bandwidth capacity of $B_{u,v}$, and the cost per traffic unit transmission of $\gamma_{u,v}$. The delay per traffic unit transmission is $d_{u,v}$. The bandwidth between SDN controller and switch $s$ is $B_{c,s}$, while $\gamma_{c,s}$ is the cost for transmission of unit of traffic. 

$C_m$ and $\gamma_m$ respectively denote the capacity and the cost per unit of storage usage of server $m$. The delay per unit-of-traffic processing of VNF type $f$ is $d^m_f$, when it is deployed on server $m$. Let $C^{l}_s$, $C^{e}_s$, and $C^{RA}_s$ be the storage capacity of respectively local, external memories, and RDMA table of the switch $s$. The cost per unit of local and external memory usage are respectively $\gamma_s^{l}$ and $\gamma_s^{e}$. The cost for in-network deployment of a VNF is higher than server-based deployment due to the high speed of a switch. Also local memory usage of a switch is more expensive than external memory due to the lower latency. The amount of control traffic per deploying VNF type $f$ in switch $s$ is $\omega_f^{s}$. The delay for processing unit of traffic of VNF type $f$, in the local memory of switch $s$ is $d_f^{s,l}$. The delay per RDMA access for switch $s$ is $d_{RDMA}^s$. $\xi_f$ is the number of RDMA accesses to fetch the required match-action entries of VNF type $f$ in RDMA table.

\vspace{-0.2cm}
\section{Optimization Formulation}
The time-horizon of the system is $T$, and each time step is $t$. Binary variables include: $x_{i,f}^{m}(t)$ indicates placement of instance $i$ of VNF type $f$ in server $m$ at time step $t$. $x_{i,f}^{s,e}(t)$ indicates placement of instance $i$ of VNF type $f$ in external memory of switch $s$. $y_{q,n}^{m,i,f}(t)$ indicates allocation of instance $i$ of VNF type $f$ which is in server $m$ to the $n-th$ VNF of service request $q$. $y_{q,n}^{s,e,i,f}(t)$ shows allocation of instance $i$ of VNF type $f$ which is in external memory of switch $s$ to the $n-th$ NVF of service request $q$. $z_{q,l}^{p}(t)$ indicates mapping of $l-th$ virtual link of request $q$ to the path $p$. $slp_{(u,v)}^p(t)$ is 1 when link $(u,v)$ is on the path $p$ at time slot $t$. Similar to external, we have variables for local memories. We follow a reconfiguration approach for placement of VNF instances on substrate nodes and steering the traffic. 

\vspace{-0.2cm}
\subsection{Objective} The configuration cost, is due to the resource usage and reconfiguration cost. The objective function is defined as (1). $\mathcal{R}(t)$ and $\mathcal{B}(t)$ are respectively the cost for IT resources and bandwidth usage at time slot $t$. $\mathcal{G}(t)$ is the reconfiguration cost due to the migration of VNFs among servers or programming the switches, and $\alpha, \beta$ are the balancing coefficients.

\vspace{-0.2cm}
{\footnotesize
\begin{equation}
\begin{aligned}
\min \sum\limits_{t\in T} \alpha.(\mathcal{R}(t) + \mathcal{B}(t)) + \beta.\mathcal{G}(t), 
\end{aligned}
\end{equation} 
}
\vspace{-0.4cm}

The IT Resource cost as calculated in (2) is the cost for deploying VNFs on the servers and local/external memories of the switches. The bandwidth cost is calculated by (3):

\vspace{-0.2cm}
{\footnotesize
\begin{equation}
\begin{aligned}
\mathcal{R}(t) = \sum\limits_{f \in F} \sum\limits_{i \in I_f} \sum\limits_{m \in M} x_{i,f}^m(t).\gamma_m.\phi_f^{m} +\\
\sum\limits_{f \in F} \sum\limits_{i \in I_f} \sum\limits_{s \in S} x_{i,f}^{s,l}(t).\gamma_s^l.\phi_f^{s} +
\sum\limits_{f \in F} \sum\limits_{i \in I_f} \sum\limits_{s \in S} x_{i,f}^{s,e}(t).\gamma_s^e.\phi_f^{s}.
\end{aligned}
\end{equation} 

\begin{equation}
\begin{aligned}
\mathcal{B}(t) = \sum\limits_{q \in Q} \sum\limits_{p} \sum\limits_{l \in L_q} \sum\limits_{(u,v) \in E} z_{q,l}^{p}(t).slp_{(u,v)}^p\lambda_q(t).\gamma_{u,v}.
\end{aligned}
\end{equation} 
}
The reconfiguration cost is the cost for the migration of VNF instances among servers ($\mathcal{M}(t)$), as well as the cost for programming the PDP swtiches ($\mathcal{N}(t)$), as below: 

\vspace{-0.3 cm}
{\footnotesize
\begin{equation}
\begin{aligned}
\mathcal{G}(t)= \mathcal{M}(t) + \mathcal{N}(t),\\ 
\end{aligned}
\end{equation}
}
where $\mathcal{M}(t)$ and $\mathcal{N}(t)$ are calculated as below:

\vspace{-0.3 cm}
{\footnotesize
\begin{equation}
\begin{aligned}
\mathcal{M}(t) = \sum\limits_{f\in{F}}
\sum\limits_{i\in{I_f}}
\sum\limits_{u\in{M}}
\sum\limits_{v\in{M}, u \neq v}
           x_{i,f}^{u}(t).x_{i,f}^{v}(t-1).\phi_f^m.\gamma_{u,v},
\end{aligned}
\end{equation} 

\begin{equation}
\begin{aligned}
\mathcal{N}(t) = \sum\limits_{f\in{F}}
\sum\limits_{i\in{I_f}}
\sum\limits_{s\in{S}}
\sum\limits_{m\in{M}}
           (x_{i,f}^{s,l}(t)+x_{i,f}^{s,e}(t)).x_{i,f}^{m}(t-1).\omega_f^{s}.\gamma_{c,s} +\\
\sum\limits_{f\in{F}} \sum\limits_{i\in{I_f}}
\sum\limits_{s\in{S}}
\sum\limits_{s'\in{S}, s \neq s'}
           (x_{i,f}^{s,l}(t)+x_{i,f}^{s,e}(t)).x_{i,f}^{s',l}(t-1).\omega_f^{s}.\gamma_{c,s} +\\
\sum\limits_{f\in{F}} \sum\limits_{i\in{I_f}}
\sum\limits_{s\in{S}}
\sum\limits_{s'\in{S}, s \neq s'}
           (x_{i,f}^{s,l}(t)+x_{i,f}^{s,e}(t)).x_{i,f}^{s',e}(t-1).\omega_f^{s}.\gamma_{c,s}
\end{aligned}
\end{equation}
}
These deployment scenarios will be involved in programming cost calculation in (6): The local/external memories programming cost with the previous deployment on the servers; and the local/external memories programming cost with the previous deployment either in local or external memory of a hosting switch different than the current host.   
\vspace{-0.3cm}
\subsection{Constraints:} 
By (7), the storage occupancy in the external memory of a switch is bounded with the existing capacity. Similarly, capacity constraints for local memories and servers are defined:

\vspace{-0.1 cm}
{\footnotesize
\begin{equation}
\begin{aligned}
\forall s \in S: \sum\limits_{f\in{F}}
\sum\limits_{i\in{I_f}}
 x_{i,f}^{s,e}(t).\phi_f^s \leq C_s^e. 
\end{aligned}
\end{equation}
}

By (8) each VNF instance will be deployed in one substrate node either a server, external/local memory of a switch:

\vspace{-0.2 cm}
{\footnotesize
\begin{equation}
\begin{aligned}
\forall f, i\in I_f: 
\sum\limits_{m\in M} 
x_{i,f}^{m}(t) + 
\sum\limits_{s\in S} 
x_{i,f}^{s,l}(t) +
\sum\limits_{s\in S} 
x_{i,f}^{s,e}(t) = 1.
\end{aligned}
\end{equation}
}
By (9), each VNF of a service request will utilize one VNF instance for the traffic processing: 

\vspace{-0.3 cm}
{\footnotesize
\begin{equation}
\begin{aligned}
\forall q, n : 
\sum\limits_{i\in{I_f}} \sum\limits_{m} 
y_{q,n}^{m,i,f}(t) + 
\sum\limits_{i\in{I_f}} \sum\limits_{s\in S} 
y_{q,n}^{s,l,i,f}(t) +
\sum\limits_{i\in{I_f}} \sum\limits_{s\in S} 
y_{q,n}^{s,e,i,f}(t) = 1.
\end{aligned}
\end{equation}
}

\vspace{-0.2 cm}
By (10), the traffic processing at each VNF instance deployed on a external memory of a switch will be bounded by corresponding capacity. Similarly, the constraints are defined for instances hosted on local memory of a switch or a server.

\vspace{-0.1 cm}
{\footnotesize
\begin{equation}
\begin{aligned}
\forall f, i \in I_f, s:
\sum\limits_{q\in Q}
\sum\limits_{n=1..|F_q|}
y_{q,n}^{s,e,i,f}(t).\lambda_q(t) \leq C_f^s. \\
\end{aligned}
\end{equation}
}

\vspace{-0.2 cm}
By (11) the link consumption meets the link capacity. By (12), for each service request, the in-flow and out-flow on each intermediate substrate node are equal, except the source and the destination nodes.

{\footnotesize
\begin{equation}
\begin{aligned}
\forall uv \in E: 
\sum\limits_{p}
\sum\limits_{q\in Q} \sum\limits_{l \in {L_q}}  z_{q,l}^{p}(t).slp_{(u,v)}^p(t).\lambda_q(t) \leq B_{uv}.
\end{aligned}
\end{equation} 

\begin{equation}
\begin{aligned}
\forall q \in Q, \forall u \in V:
\sum\limits_{v\in nb(u)} 
\sum\limits_{p} 
\sum\limits_{l \in L_q} 
z_{q,l}^{p}(t).slp_{(u,v)}^p(t) -\\
\sum\limits_{v\in nb(u)} 
\sum\limits_{p} 
\sum\limits_{l \in L_q} 
z_{q,l}^{p}(t).slp_{(v,u)}^p(t) = 
\left\{
\begin{matrix}
1 &  u = o_q \\
-1 & u = t_q \\
0 & otherwise
\end{matrix}
\right.
\end{aligned}
\end{equation}
}

The traffic processing delay of each service request $q$ i.e., ${D}_q(t)$ should meet the deadline i.e., $\forall q \in Q: \mathcal{D}_q(t) < d_q(t).$

\vspace{-0.1cm}
\subsection{Delay Calculation} 
\noindent \textbf{Intra-Server Migration Delay.} As the VNF migrations among servers can be performed in parallel, the maximum migration time specifies the migration delay:

\vspace{-0.2 cm}
{\footnotesize
\begin{equation}
\begin{aligned}
\mathcal{\psi^M}_q(t) = 
\max_{n}
\sum\limits_{f\in{F}}
\sum\limits_{i\in{I_f}}
\sum\limits_{u\in{M}}
\sum\limits_{v\in{M}, u \neq v}
           y_{q,n}^{u,i,f}(t).y_{q,n}^{v,i,f}(t-1).\frac{\phi_f^m}{B_{u,v}}.
\end{aligned}
\end{equation}
}

\noindent \textbf{SDN Controller Programming Delay.} The maximum communication time with SDN controller to program switches to process the VNFs, specifies the programming delay as calculated by (14). Under any reconfiguration scenario described in (6), the delay is calculated as the ratio of the traffic required for programming the switches and the bandwidth between SDN controller and switches.    

\vspace{-0.6cm}
{\footnotesize
\begin{equation}
\begin{aligned}
\mathcal{\psi^N}_q(t) = \max_{n}\\
\sum\limits_{f\in{F}}
\sum\limits_{i\in{I_f}}
\sum\limits_{s\in{S}}
\sum\limits_{m\in{M}}
           (y_{q,n}^{s,l,i,f}(t)+y_{q,n}^{s,e,i,f}(t)).y_{q,n}^{m,i,f}(t-1).\frac{\omega_f^{s}}{B_{c,s}} +\\
\sum\limits_{f\in{F}} \sum\limits_{i\in{I_f}}
\sum\limits_{s\in{S}}
\sum\limits_{s'\in{S}, s \neq s'}
           (y_{q,n}^{s,l,i,f}(t)+y_{q,n}^{s,e,i,f}(t)).y_{q,n}^{s',l,i,f}(t-1).\frac{\omega_f^{s}}{B_{c,s}} +\\
\sum\limits_{f\in{F}} \sum\limits_{i\in{I_f}}
\sum\limits_{s\in{S}}
\sum\limits_{s'\in{S}, s \neq s'}
           (y_{q,n}^{s,l,i,f}(t)+y_{q,n}^{s,e,i,f}(t)).y_{q,n}^{s',e,i,f}(t-1).\frac{\omega_f^{s}}{B_{c,s}}.
\end{aligned}
\end{equation} 
}
\textbf{RDMA Delay.} (15) calculates the delay of processing the unit of traffic in external memory. The function flow table entries are fetched $\xi_{f}$ times and are applied on the traffic (Fig. 1): 

\vspace{-0.2cm}
{\footnotesize
\begin{equation}
\begin{aligned}
 d_f^{s,e} =  \xi_{f}.(d_f^{s,l} + d_{RDMA}^s)=[\frac{\phi_f^s}{C^{RA}_s}].(d_f^{s,l} + d_{RDMA}^s) 
\end{aligned}
\end{equation} 
}

\noindent \textbf{Traffic Processing.} Delay includes the reconfiguration delay ($\mathcal{\psi}_q(t)$), the VNF processing delay ($\mathcal{S}_q(t)$), and the transmission delay ($\mathcal{T}_q(t)$), as defined below:

\vspace{-0.2cm}
{\footnotesize
\begin{equation}
\begin{aligned}
 \mathcal{D}_q(t) = \mathcal{\psi}_q(t) + \mathcal{S}_q(t) + \mathcal{T}_q(t).        
\end{aligned}
\end{equation} 
}
The reconfiguration delay for SFC $q$ is due to the migration of required VNF instances among servers, as well as the delay for programming the swtiches to process the required VNFs. As both migration of VNFs and reprogramming of switches can be performed in parallel, the maximum of the caused delays specifies the reconfiguration delay: 

\vspace{-0.2cm}
{\footnotesize
\begin{equation}
\begin{aligned}
\mathcal{\psi}_q(t)= \max(\mathcal{\psi^M}_q(t), \mathcal{\psi^N}_q(t)).\\ 
\end{aligned}
\end{equation}
}

Processing delay in (16) is calculated as sum of the the traffic processing delay in the VNFs of the chain hosted either in the servers, or local/external memories of switches:

\vspace{-0.2 cm}
{\footnotesize
\begin{equation}
\begin{aligned}
 \mathcal{S}_q(t) =  
 \sum\limits_{n=1}^{|F_q|} \lambda_q.[ 
 \sum\limits_{m \in M} \sum\limits_{f \in F} \sum\limits_{i \in I_f} y_{q,n}^{m,i,f}(t).d_m^f +\\     
  \sum\limits_{s \in S} \sum\limits_{f \in F} \sum\limits_{i \in I_f} y_{q,n}^{s,l,i,f}(t).d_f^{s,l} + y_{q,n}^{s,e,i,f}(t).d_f^{s,e}], 
\end{aligned}
\end{equation} 
}

\vspace{-0.2cm}
Traffic transmission delay in (16) is calculated as below:
\vspace{-0.2cm}
{\footnotesize
\begin{equation}
\begin{aligned}
 \mathcal{T}_q(t) =  
 \sum\limits_{l=1}^{|L_q|} \lambda_q(t). 
 \sum\limits_{p}
 \sum\limits_{(u,v) \in E} z_{q,l}^{p}(t).slp_{(u,v)}^p.d_{u,v} 
\end{aligned}
\end{equation} 
}

\section{DRL Based SFCs Reconfiguration}
The basic problem of SFC provisioning is NP-hard \cite{dong2020application} and MDP/RL as advocated in solutions for dynamic network situations, can be used for formulation since: (i) Function (1) is the sum of the function value at $t -1$ and the cost at $t$. Thereby, having the memory-less property; (ii) Considering the parameters determining the current state e.g., resources storage, bandwidth status, association of VNF instances to nodes, every action that is performed by the agent ends to a new state transition, that only depends on the current state; (iii) The function (1) is in the form of accumulated rewards. Through an iterative process of observing the state, choosing an action, and receiving a reward, the state-action Q-values are estimated by Bellman equation \cite{mnih2015human}. The high dimensions of the states/actions and the dynamicity in state transitions makes observing all states and actions in training impossible, thereby inefficiency of conventional RL. To deal with this problem, we adapt DRL \cite{mnih2015human}, that generalizes experienced states/actions to non-observed ones through a neural network-based approximation of Q-values.

\vspace{-0.2cm}
\subsection{MDP Elements}
\noindent \textbf{State}: The state features at time step $t$, are as below:\\
(i) The remained storage capacity of servers: $rc(t)=[rc_1(t) ... rc_{|M|}(t)]$.\\
(ii) The remained storage of external memories in switches: $rc_s^e(t)=[rc_1^e(t)...rc_{|S|}^e(t)]$. Similarly $rc_s^l(t)$ is defined.\\ 
(iii) The remained capacity of VNF instances: matrix $rc_f(t)$, at which the entry at row $i$, and column $j$ is the remaining capacity of instance $i$ of VNF type $j$.\\
(iv) Let the sum of adjacent links' remained bandwidth of substrate node $v$ be $rb_v(t)$. The vector of {\footnotesize$rb(t)=[rb_1(t),...rb_{|M|+|S|}(t)]$} are in state features.\\
(v) The current placement of VNF instances on servers, local and external memories of switches: {\footnotesize$\{x_{i,f}^m(t), x_{i,f}^{s,l}(t), x_{i,f}^{s,e}(t)\}$}

\noindent \textbf{Action:} Actions are the placement of VNF instances over servers, local and external memories of switches for the next time step. (values of {\footnotesize$x_{i,f}^m$, $ x_{i,f}^{s,l}$, $ x_{i,f}^{s,e}$} for the next time step). 

\noindent \textbf{Reward:} To ensure the validity of deployment, the reward will have 0 value if the capacity constraints 7, 10 be violated. Otherwise, to ensure high acceptance ratio, while optimizing the objective function the reward is calculated as (20). $acr(t)$ is the ratio of requests that meet their deadline. $\alpha_r$ and $ \beta_r$ are the weights defining the priority of acceptance ratio and cost.  

\vspace{-0.2cm}
{\footnotesize
\begin{equation}
    \begin{aligned}
    \mathbb{R}(s(t),a(t))= \alpha_r.acr(t) + \frac{\beta_r}{\alpha.(\mathcal{R}(t) + \mathcal{B}(t)) + \beta.\mathcal{G}(t)}
    \end{aligned}
\end{equation}
}

\begin{figure*}[!h]
\begin{center}
    \begin{minipage}{8.4cm}
    \begin{center}
        \includegraphics[width=8.5cm, height=3.9cm]{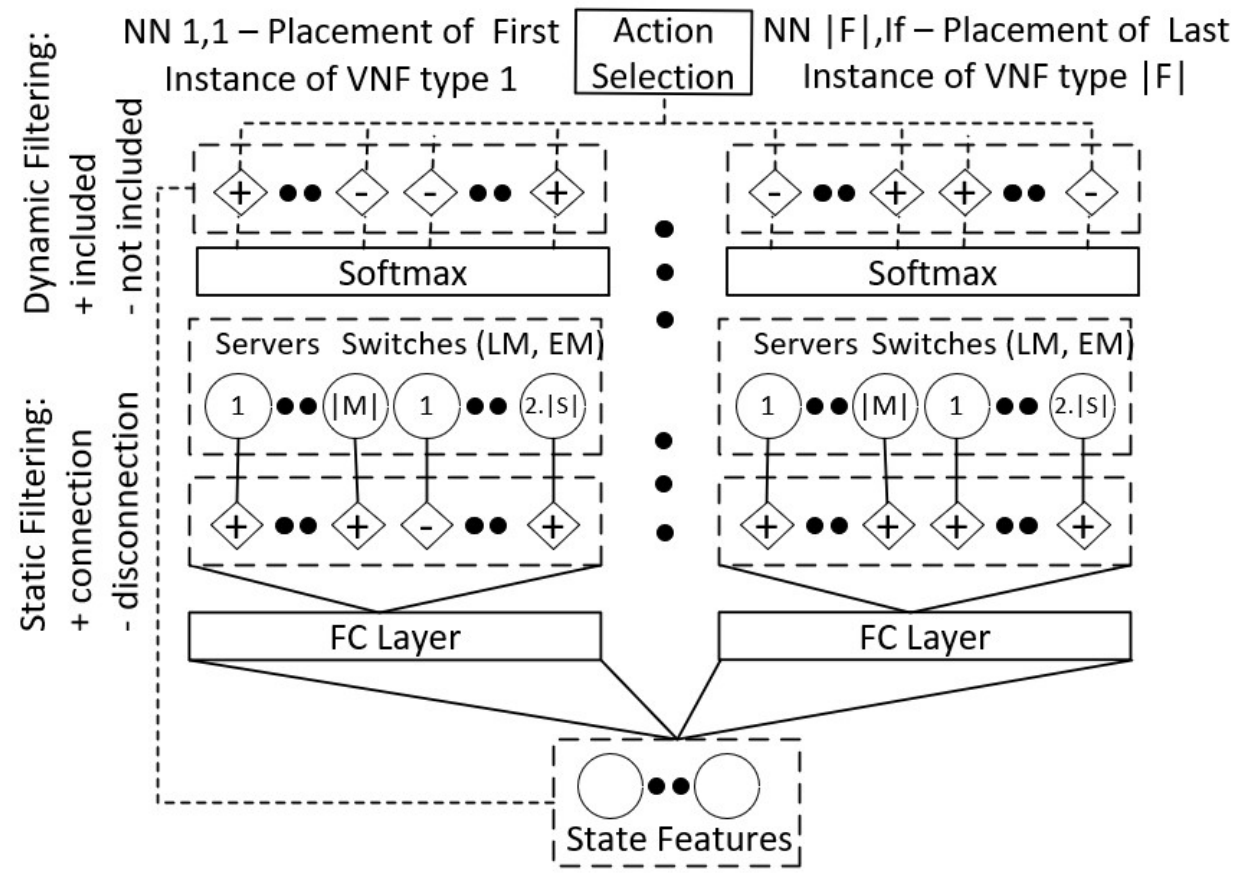}
        \label{fig:NN}
        \vspace{-0.6cm}
        \caption{Policy networks. For each switch there are 2 neurons indicating Local Memory (LM) and External Memory (EM).}
    \end{center}
    \end{minipage}
    \begin{minipage}{2.8cm}
    \begin{center}
        \includegraphics[width=2.8cm, height=3.9cm]{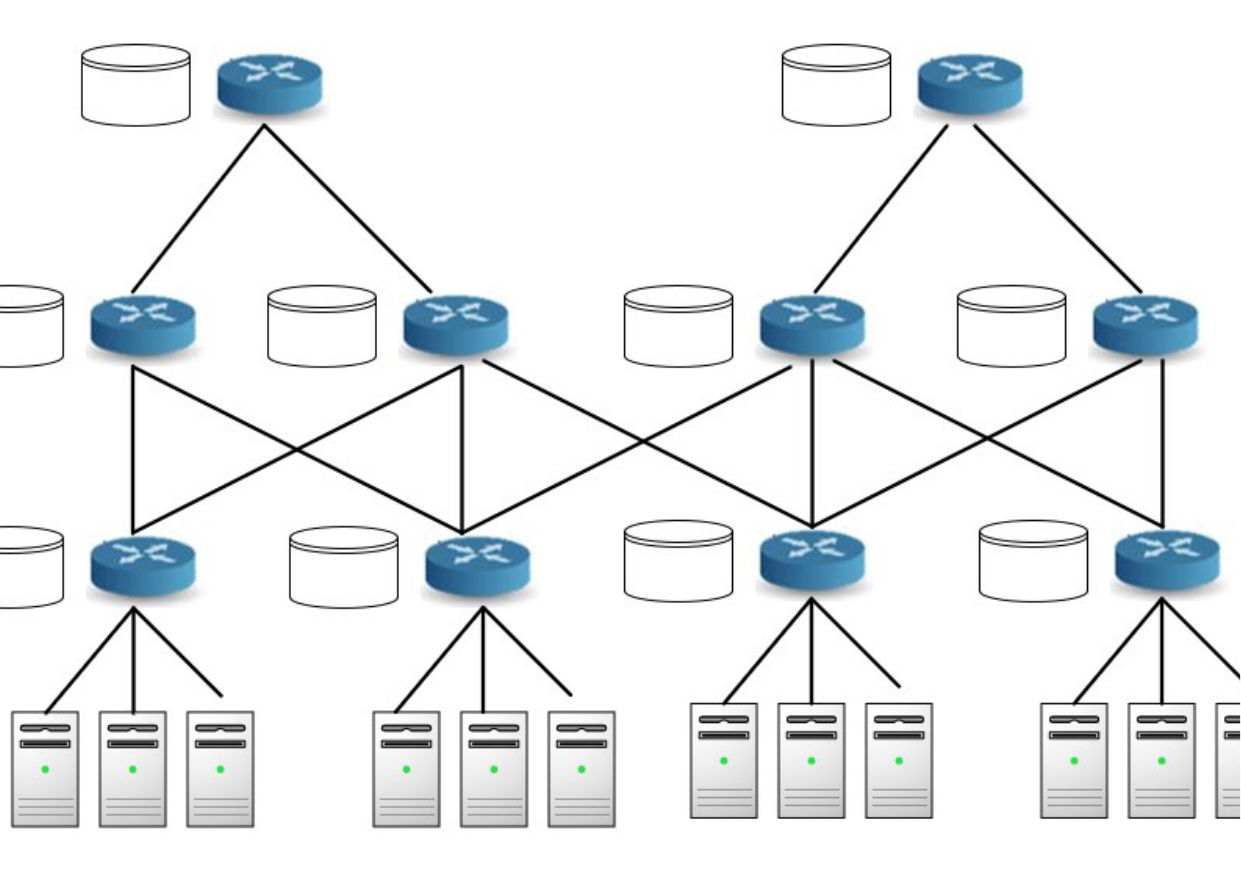}
        \label{fig:conf1}
        \vspace{-0.6cm}
        \caption{Topology of fat tree.}
        \end{center}
    \end{minipage}
    \begin{minipage}{5.6cm}
    \begin{center}
        \includegraphics[width=5.6cm, height=3.9cm]{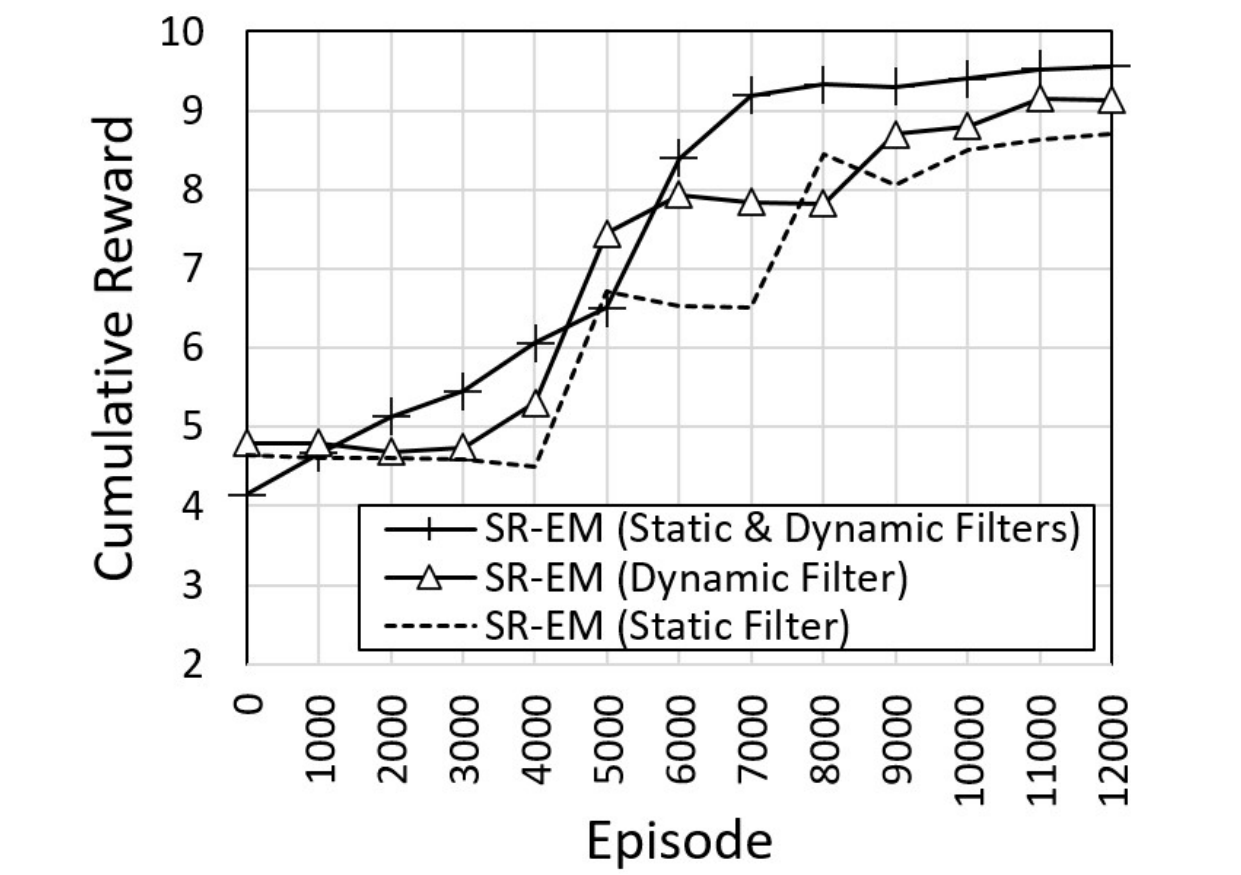}
        \label{fig:conf1}
        \vspace{-0.4cm}
        \caption{Reward variation within training.}
    \end{center}
    \end{minipage}
\end{center}
\vspace{-0.8cm}
\end{figure*} 
\textit{By optimizing cumulative reward in lifetime of SFCs, the in-network or server-based processing will automatically be adjusted considering the dynamicity in the SFCs traffic rates/QoS variation and available processing (See Fig. 2 and 6.c).}
\vspace{-0.3cm}
\subsection{Policy Network and Training}
Fig. 3 illustrates the policy networks, according which the decision policy is derived by training them. There are $|F|.|I_f|$ Neural Networks (NNs), each representing the placement policy for one VNF instance. The input neurons of each NN are the state features. In an abstract view, there is a Fully-Connected layer, with Softmax activation function. The output neurons of each NN, with size $|M|+2|S|$ indicate the probability of deployment of the associated VNF instance on the servers, the local and the external memories of switches. Since the incorporation of external memories makes deployment scenarios complex, it needs more exploration and slows down the convergence. Partially connected layers are emulated by adding the below filtering layers in training:
\vspace{-0.05 cm}
\begin{itemize}
    \item \textbf{Static Filtering:} There is connectivity between state features and a neuron if the storage capacity of the associated server or switch local memory or external memory can accommodate the storage demand of VNF. 
    \item \textbf{Dynamic Filtering:} It will be applied dynamically at every time interval according to the state feature values. For a migration decision, only the servers or local/external memories of switches are explored that their current remaining storage can accommodate the VNF instance (constraint 7). The sign $+$ in Fig. 3 activates the output neuron in exploration.      
\end{itemize}
\vspace{-0.1 cm}
The filters block the exploration of unfeasible policies and reduce the number of weights, to speed up convergence. Algorithm 1 is the implementation. Output neurons are labeled so that in the case that they are deactivated \textit{to be distinguished for VNF placement}, by their node type ('s' for switch, and 'm' for server), index, and memory type ('EM' and 'LM'). In an episode, the networks are trained to automatically adjust in-network and server based processing within the lifetime of services for an initial configuration selected randomly from a set of configurations in the train set. Training is done through three steps performed at every time interval of an episode: 

\noindent (i) \textit{Action Selection:} By $\epsilon-$greedy policy, with a $\epsilon$ probability, a random placement of VNFs over servers, local/external memories of switches is selected. Otherwise, the current state features are given as input to the NNs. At each NN, the server or local/external memory of a switch, with the highest probability at output layer, will be selected for the placement of the associated VNF instance. 

\noindent (ii) \textit{Link placement:} A Dijkstra-based shortest path method at which the load of links are the weights for load-balancing, is used to map the virtual links to the physical routes. When a VNF instance is migrated between servers or programmed on a switch, the same service requests as previous time step are assigned to the VNF instance (ensuring constrain 9). 

\noindent (iii) \textit{Updating the weights:} After the VNF instances placement and link mapping, reward is calculated for the new configuration by (20), accordingly the NNs' weights are updated by Gradient Descent method and Bellman equation \cite{mnih2015human}.  

\vspace{-0.2cm}
\setlength{\textfloatsep}{0 cm}
\begin{algorithm}
\footnotesize
\caption{Static and Dynamic Filtering}
\textbf{Static Filter}

\For{$NN _{i,f}$ associated with instance $i$ of VNF type $f$}
{
    $N_{i,f} \gets$ 0
    $lables \gets \{\}$

    \{Similarly for LM\}\\
    \For{each switch $s$}
    {
        \If{$\phi^{s}_f < C^{e}_{s}$}
        {
             $N_{i,f} \gets N_{i,f} + $ 1\\ 
             $labels \gets lables \cup \{<'s',~s,~'EM'>\}$ 
        }
    
    }

    \For{each server $m$}
    {
        \If{$\phi^{s}_f < C^{m}$}
        {
             $N_{i,f} \gets N_{i,f} + $ 1 
             $labels \gets lables \cup \{<'m',~m,~'m'>\}$
        }
    
    }
}
Deploy each $NN _{i,f}$ with hidden neurons count of $N_{i,f}$ \\
\textbf{Dynamic Filter}\\
$Actions \gets \{\}$\\
\For{each neuron in output layer of $NN_{i,f}$}
{  
    \If {at previous interval, instance $i$ of $f$ was not placed on the associated node (defined by the neuron label)}{
        \{Similarly for LM and server\}\\
        \If {label of neuron is 's' and is of type EM}
        {
            \If {$rc^{e}_{s}(t) > \phi^{s}_f$}
            {
                $Actions \gets Actions \cup \{index~in~label~ i.e., s\}$
            }
            
        }
            
    }
}
\vspace{-0.1cm} 
\end{algorithm}

\section{Evaluation Results}
The network topology is a fat-tree with 10 switches and 12 servers (Fig. 4). The storage capacity of the servers, \textbf{L}ocal \textbf{M}emories (LM) and \textbf{E}xternal \textbf{M}emories (EM) of switches are random in the ranges of $[10, 40]$ GB, $[10, 35]$ MB, and $[100, 500]$ MB, respectively \cite{kianpisheh2022survey}. The RDMA tables' sizes are the same as LMs. The links bandwidth are within $[30, 50]$ Gbps. The bandwidth between SDN controller and switches is 50 Gbps. The propagation delay of the links are in $[0.5, 1]$ ms. There are 4 VNF types, each with 2 instances. The VNFs storage demands are in $[10, 50]$ MB. The data for programming a VNF in a switch is the same as its size \cite{aghaaliakbari2023architecture}. The traffic processing capacity of VNFs are in $[25, 40]$ Mbps and 100 Gbps, when they are deployed respectively in the servers and switches. The delay per MB processing of chains traffic is in $[0.1, 0.3]$ ms for servers \cite{mouradian2019application}, and $[0.01, 0.03]$ ms for switches. The RDMA delay is in the range of $[50, 150]$ ns per access. 

The cost per GB memory consumption at servers are in the ranges of $[0.002, 0.003]$ of unit of currency, while it is in $[0.0021, 0.0024]$ per MB local memory storage utilization in switches. The costs are reduced per MB external memory usage in switches with amount in the range of $[0.0002, 0.0004]$ of unit of currency, due to RDMA delay. The cost per GB transfer is in proportion with BW and in the range of $[0.8, 1]$.

SFCs include up to 4 random subset of VNF types. Sources and destinations are random switches. For each SFC, the traffic flow is generated from 10 AM to 22 PM, with a pattern of traffic higher during the working hours till 4 PM, in the range of $[15, 35]$ Mbps, and lower after 4 PM, in the range of $[200, 500]$ Kbps (according to the demand of Web service, VoIP, and online Gaming \cite{pham2017traffic}). The peak SFC-traffic at working-hours is consistent with the analysis in \cite{troia2019reinforcement}. 

The processing deadlines are in the ranges of $[3, 8]$ and $[8, 12]$ ms, respectively for non-working-hours, and working hours. Train and test set consists of 1200 and 300 instances, respectively. Each instance has placed 90 SFCs in the network. Initially, VNF instances are deployed on the servers with minimum cost that can provide the required storage. Then, a VNF in a request is mapped to a random VNF instance that can accommodate the chain traffic flow. For each train/test instance, VNFs deployments are reconfigured every two hours, due to the SFC's traffic load/deadline variations. More explorations at early iterations of DRL has been used, while the exploitation increases up to 98\%-greedy selection at the last episode. The values of 0.2 and 0.3 respectively for GD optimization learning rate and discount rate, operated efficient. We have set: $\alpha_r = 0.8$, $\beta_r = 0.2$, $\alpha = 1$, $\beta = 1$. Each episode is assigned with a random instance in the train set. 

As shown in Fig. 5, inclusion of only dynamic filters achieves more reward than having only static filters. Since, dynamic filtering is more general than static filtering to prune unfeasible placements explorations during training. In the hybrid case, by exploiting reduction of weights count by static filtering and the capability of pruning unfeasible exploration of dynamic filtering the gained rewards and convergence status is enhanced, as it is stable after episode 7000. Benchmarks are: a) RANDOM: it migrates the VNF instances randomly among servers, LM and EM of switches which can accommodate the instance; b) DDQN-CM \cite{afrasiabi2022reinforcement}: it migrates VNFs among servers to minimize the delays and migration cost. The state of MDP indicates the placement of VNFs in servers. The actions are possible servers for VNF migration. The reward is the accumulation of chains' delay and migration cost. DDQN (a DRL method) with the same learning parameters in \cite{afrasiabi2022reinforcement}, is trained 12000 episodes for reconfiguration; c) LAG \cite{li2022qos}: it considers an environment with servers and switches with LMs. The heuristic in \cite{li2022qos} is applied, for SFC requests reconfiguration to minimize the total cost (IT/bandwdth/reconfiguration cost). A SFC is divided into several segments, each representing the connection between two adjacent VNFs. Duplication of Substrate Network (SN) topology, called as LAGs and represented with graphs, are utilized to map each segment to SN. In reconfiguration, for each LAG, the servers/switches that can not provide the latency ($d_q/|F_q|$) or bandwidth demands are removed. For each VNF at a LAG, a node that can host the VNF with the smallest cost is selected. The VNFs among LAGs are connected by the shortest path. 

\begin{figure}[t]
\begin{center}
    \subfigure{
    \includegraphics[width=6.3cm, height=2.7cm]{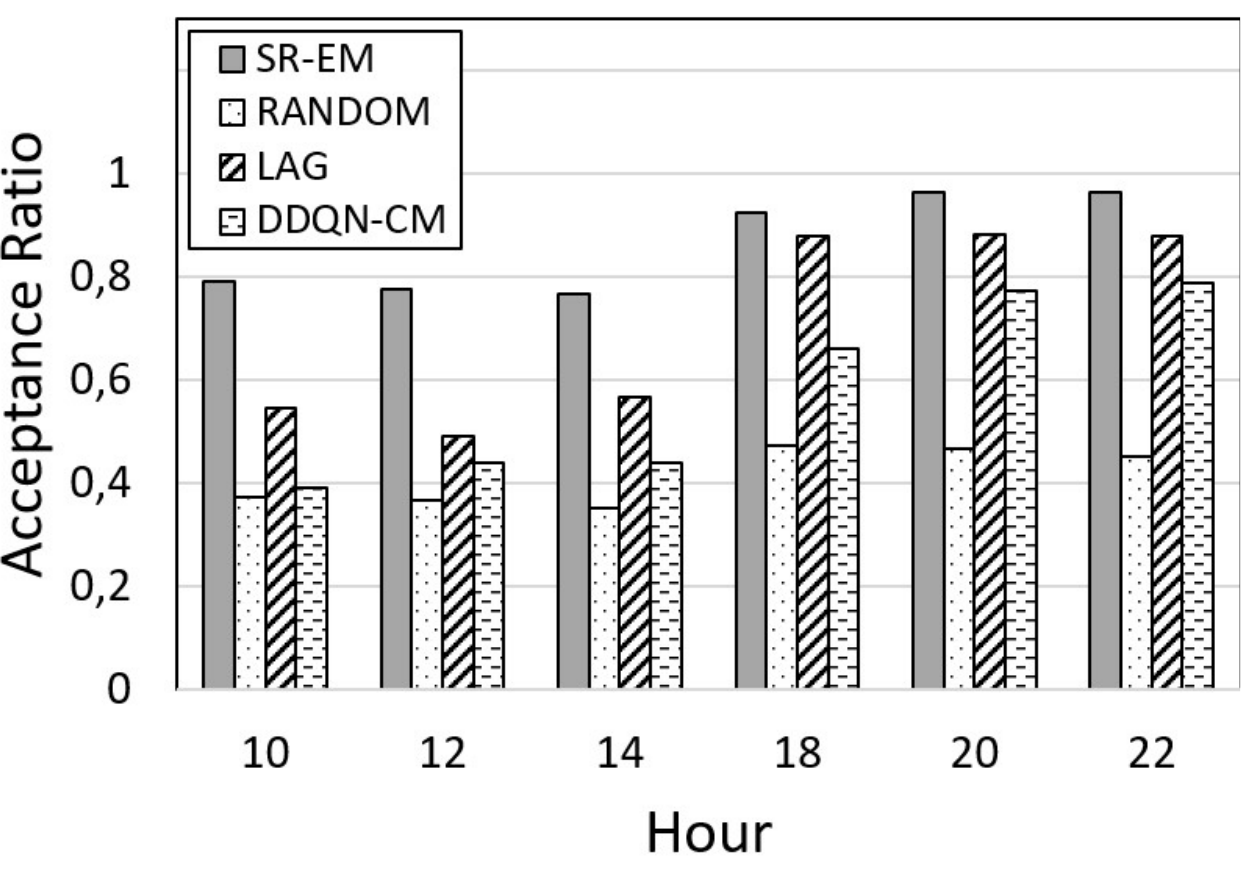}
    \label{fig:acc}
    }    
    \vspace{-0.3cm}
    \subfigure{
    \includegraphics[width=6.3cm, height=2.7 cm]{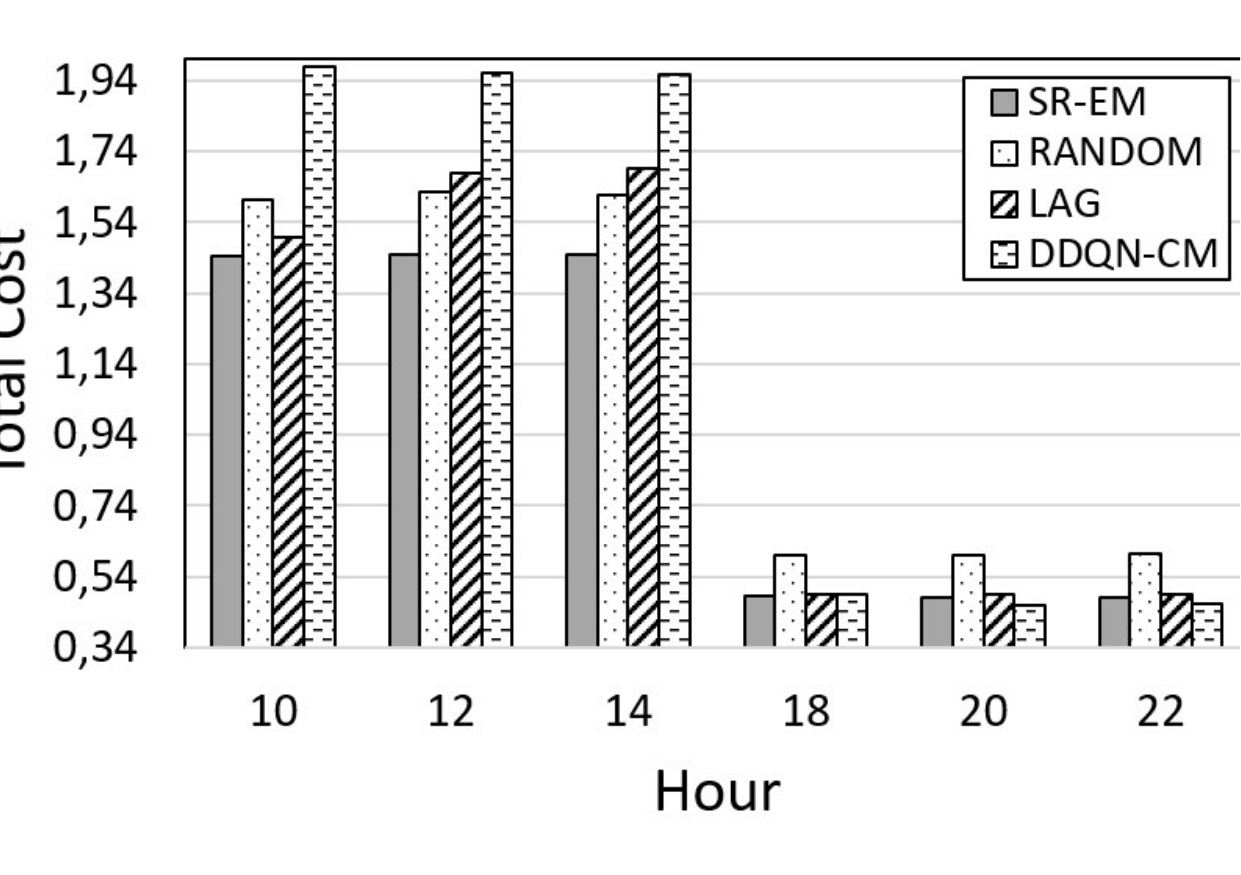}
     \label{fig:cost}
    }    
    \vspace{-0.3cm}
    \subfigure{
    \includegraphics[width=6.3cm, height=2.7 cm]{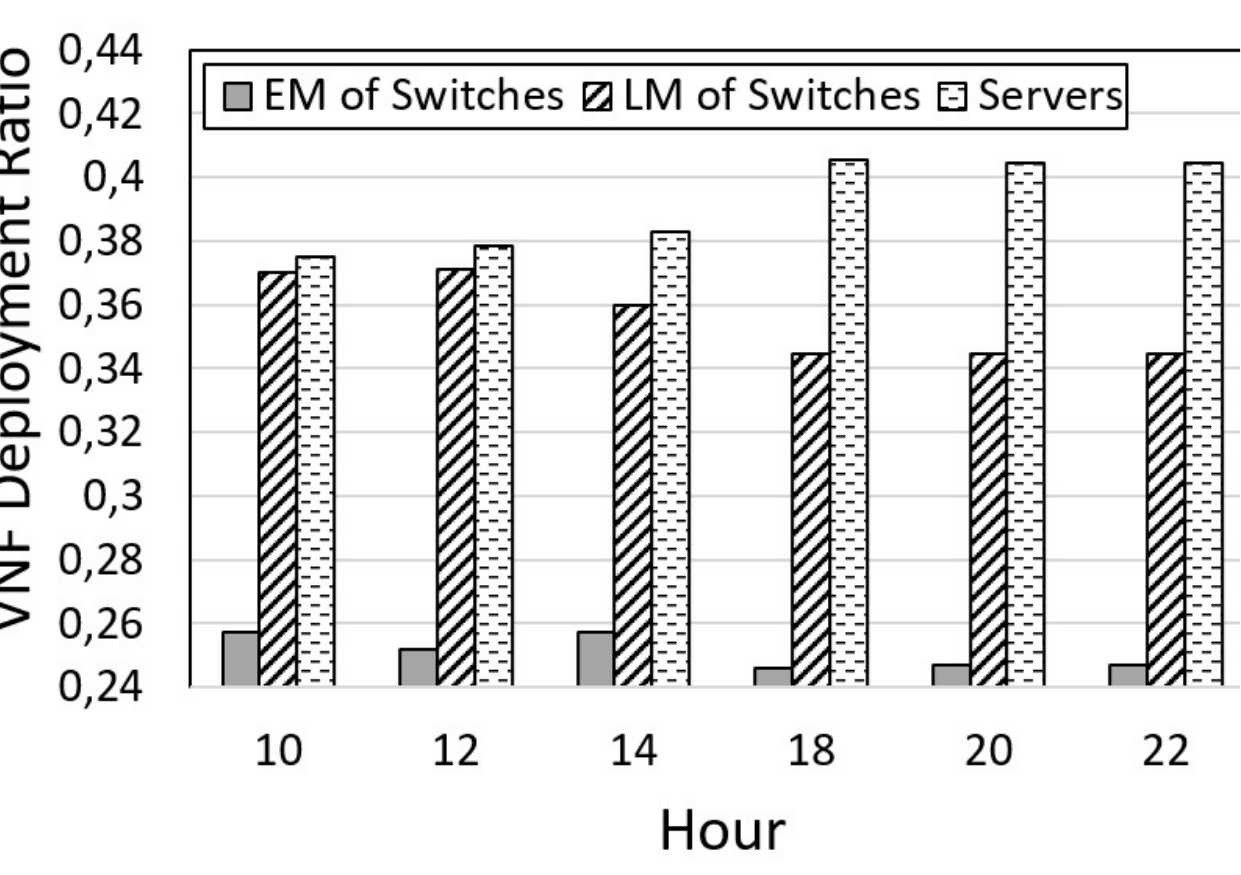}
    \label{fig:dep}
    }
    \vspace{-0.2cm}
    \caption{Evaluation results. (a) Mean acceptance ratio. (b) Mean total cost. (c) In-network and server based processing adaption in SR-EM.}   
    \vspace{-0.3cm}
\label{fig:comp}
\end{center}
\end{figure}

Fig. \ref{fig:comp}.a shows the acceptance ratio for the test set. In all methods, the acceptance ratio is lower in working hours due to the higher traffic load. DDQN-CM has accepted lower requests than LAG. Since LAG exploits high-speed processing of VNFs in LMs of switches. SR-EM has increased acceptance ratio up to 29\% and 40\% in comparison with LAG and DDQN-CM, due to enhancing the in-network processing of VNFs through exploiting both LMs and EMs for VNFs deployment. Low performance of RANDOM illustrates the efficiency of DRL in SR-EM for optimal configuration.   

Fig. \ref{fig:comp}.b shows the total cost. In working-hours cost is higher than non-working-hours due to high bandwidth usage. In working hours with high traffic load, the dominant cost is the bandwidth cost. In comparison with DDQN-CM, INC reduces bandwidth cost, and accordingly the total cost, by avoiding detouring of data to the distant servers (sources/destinations are closer to the switches.) SR-EM that adds EM as another opportunity for INC has the lowest cost (up to 0.53 unit of currency decrease). Random exploitation of EM is not as efficient as DRL-based placement in SR-EM. In the non-working-hours, the IT-cost is the dominant cost. LAG that utilizes more expensive resources than the servers needs more cost than DDQN-CM (up to 0.03 unit of currency increment). In comparison with LAG, SR-EM has slightly reduced cost, almost 0.01 unit of currency, due to a cheaper resource price of EMs. 

Fig. \ref{fig:comp}.c shows automatic adoption of in-network and server based processing in various traffic loads in SR-EM. For non-working-hours with light traffic, servers that are cheaper are utilized more than LMs of switches. LMs of switches with less delay, is used more than EMs, to meet the low latency processing requirements. The usage of EMs has complemented LMs to enhance acceptance ratio. It is used more in working-hours, since large-size VNF instances deployed over the servers may not have enough capacity in high traffic loads and EMs can host those VNF instances.

\vspace{-0.3cm}
\section{Conclusion and Future Work}
This paper studies service function chain reconfiguration problem in NFV environment while exploiting the external memories of PDP switches and adoptable with dynamic network and chains' traffic characteristics. The reconfiguration is modeled as an optimization that minimizes total deployment and reconfiguration cost while meeting the time and resources constraints. A converged-enhanced DRL based method is proposed to solve the optimization while considering dynamic nature of the network and SFCs traffic. Results illustrates the improvement in acceptance ratio and total cost. Assessing the efficiency of INC in power saving is a future work.

\bibliographystyle{IEEEtran}
\bibliography{ref.bib}
\end{document}